\documentclass[10pt,aps,prl,twocolumn,floatfix]{revtex4-2}
\usepackage{epsfig}
\usepackage{graphicx}
\usepackage{dcolumn}
\usepackage{amsthm,amsmath,amsfonts}
\usepackage{mathtools}
\usepackage{comment}
\usepackage{xcolor}
\usepackage{orcidlink}
\usepackage{rotating}
\usepackage{cancel}
\usepackage[normalem]{ulem}
\usepackage{bm}
\usepackage{latexsym}
\usepackage{verbatim}
\usepackage{comment}
\usepackage{stmaryrd}
\usepackage{color}
\usepackage{amsmath,amssymb,graphicx}
\usepackage{braket}
\usepackage{booktabs}
\usepackage{lineno}
\usepackage{bm}

\begin{document}
\title{Applications of quantum annealing to magnetic dipole hyperfine structure constants: First results beyond energies for atoms}

\author{$^{a,b}$Boni Paul\orcidlink{0009-0007-8355-6571}}
\email{boni.paul.71@tcgcrest.org}

\author{$^a$Subimal Deb\orcidlink{0000-0001-7697-1266}}
\email{subimal.deb@tcgcrest.org}

\author{$^c$Per Jönsson\orcidlink{0000-0001-6818-9637}}
\email{per.jonsson@mau.se}

\author{$^c$Jörgen Ekman}
\email{jorgen.ekman@mau.se}

\author{$^a$Bhanu Pratap Das\orcidlink{0000-0002-3766-6979  }}
\email{Bhanu.das@tcgcrest.org}

\affiliation{$^a$Centre for Quantum Engineering Research and Education, TCG Centres for Research and Education in Science and Technology, Sector V, Salt Lake, Kolkata 700091, India}
\affiliation{$^b$Department of Physics, Indian Institute of Technology Tirupati, Yerpedu, Andhra Pradesh 517619, India.}

\affiliation{$^c$Department of Materials Science and Applied Mathematics, Malmö University, SE-20506 Malmö, Sweden}


\begin{abstract}
We report the first results of the magnetic dipole hyperfine structure (HFS) constants of neutral $\mathrm{Li}$, Li-like $\mathrm{Be}$, neutral $\mathrm{Na}$, and Na-like $\mathrm{Mg}$ using a modified version of the Quantum Annealer Eigensolver (QAE) algorithm on D-Wave's quantum hardware. The results are benchmarked against relativistic configuration interaction with multiconfiguration Dirac Hartree-Fock (MCDHF)  calculations using the General-purpose Relativistic Atomic Structure Package (GRASP), and simulated annealing. In our modified QAE, a zooming-and-sigma-annealing approach with a floating-point encoding scheme is adopted to estimate the ground-state eigenvalue and eigenvector of the relativistic Dirac-Coulomb Hamiltonian matrices ($H_{\mathrm{DC}}$) constructed from 11 or fewer configuration state functions (CSFs). For calculations with extended correlation orbital sets, we applied a CSF truncation scheme, retaining only  CSFs (up to 12) that make significant contributions to the ground-state wavefunction. Our modified QAE precision is kept limited to three decimal places (up to 10 qubits). Hardware demonstrations on the D-Wave quantum processing unit (QPU) yielded results that were completely consistent with GRASP (at the chosen precision) in determining the magnetic dipole HFS constants, with accuracy varying across systems and $H_{\mathrm{DC}}$ matrix dimensions.
\end{abstract}

\maketitle 

Quantum annealing (QA) \cite{PhysRevE.58.5355, Hauke_2020} is a meta-heuristic optimization method that utilizes quantum effects such as tunneling and superposition to find the ground state of a problem Hamiltonian. It is a specialized method related to adiabatic quantum computing \cite{10.1098/rsta.2021.0419, Yarkoni_2022, Crosson2021}. A recent achievement in quantum annealing shows that it can outperform classical simulations for some specific state-of-the-art many-body problems, despite current quantum hardware limitations \cite{doi:10.1126/science.ado6285}.   The Quantum Annealer Eigensolver (QAE) \cite{Teplukhin2019} is a quantum–classical hybrid algorithm that solves eigenvalue problems by minimising a suitable objective function. It has demonstrated applicability to diverse problems, including the fine structure splitting calculations which involves the computation of the difference between excited and ground state energies in highly charged ions \cite{kumar2024computation}, complex eigenvalue problems \cite{D0CP04272B}, strong correlation effects such as avoided crossings in $H_4$ molecule \cite{Zade2025}, molecular electronic state energies  \cite{Teplukhin2020, Teplukhin2021}, molecular vibrational spectra  \cite{Teplukhin2019}, particle physics simulations \cite{PhysRevA.106.052605}, lattice gauge theories  \cite{PhysRevD.104.034501}, and permanent electric dipole moment calculations for molecules\cite{sahoo2025applicationquantumannealingcomputation}. The versatility of the QAE algorithm, its robustness against quantum hardware errors and its limited current applications for accurate physical property computations, motivate further explorations of various atomic studies using this algorithm. One such important case is the hyperfine interaction \cite{PhysRev.97.380, armstrong_1971,Das1973}, which plays a fundamental role across diverse areas of physics \cite{Kaur_2020}. Current timekeeping standards are based on hyperfine transitions \cite{Essen1955,Ramsey1983}, and state-of-the-art optical clocks also exploit highly forbidden hyperfine-induced clock transitions \cite{RevModPhys.87.637,Beloy2021}. Beyond metrology, accurate hyperfine-structure determination is essential for interpreting the cosmological 21-cm hydrogen line \cite{Vrbanec2019}, advancing cold-atom physics \cite{doi:10.1126/science.269.5221.198,PhysRevLett.75.3969} and determination of nuclear structure related to magnetic dipole and higher order nuclear moments \cite{PhysRevLett.48.1330,Das1973}. These diverse applications underscore the importance of developing accurate quantum algorithms as alternatives to their classical counterparts for addressing problems like hyperfine structure calculations.

Accurate calculations of atomic properties beyond ground-state energies present significant challenges in capturing electron correlation effects, even when computing the same property across different systems \cite{VajedSamii1982} or different properties for the same system \cite{PhysRevD.103.L111303}. The magnitude and importance of correlation contributions can vary substantially depending on the specific observable under consideration \cite{PhysRevD.103.L111303}. Computing hyperfine structure (HFS) constants exemplifies these challenges, as such calculations impose substantially more stringent requirements than energy determinations on quantum hardware \cite{PhysRevA.110.062620}. The near-nuclear region having pronounced relativistic and non-relativistic effects, demands a precise determination of wavefunctions \cite{PhysRevA.72.032507}. This motivates our investigation into the capability of quantum annealers to capture these subtle yet crucial many-body effects.  As a first step toward practical computations of fundamental atomic properties, we employ the QAE algorithm on a D-wave quantum annealer to compute the magnetic dipole HFS constants for four atomic systems: neutral $\mathrm{Li}$, Li-like $\mathrm{Be}$, neutral $\mathrm{Na}$, and Na-like $\mathrm{Mg}$. We focus on the lowest angular symmetry states with total electronic angular momentum $J = 1/2$ and even parity. While comparative calculations of HFS constants have been performed on gate-based quantum computers for Li and Li-like systems \cite{PhysRevA.110.062620}, to the best of our knowledge, this work is the first calculation of HFS constants on a quantum annealer, thereby extending the QAE framework beyond previously studied ground-state atomic energies. We compare our results with classical calculations using the configuration interaction (CI) method with multiconfiguration Dirac-Hartree-Fock (MCDHF) orbitals using the General-Purpose Relativistic Atomic Structure Package (GRASP) \cite{atoms11010007,FroeseFischer2019GRASP}  and simulated annealing (SA)\cite{doi:10.1126/science.220.4598.671}.

The remainder of this Letter is organized as follows. We first outline the theory of magnetic-dipole HFS constants and give an overview of the computational workflow. We then describe our modified QAE methodology for computing ground-state energy and wavefunction, from which HFS constants are extracted. Next, we present the results obtained from the D-Wave quantum annealer, comparing them with GRASP and SA. Finally, we summarize our findings and discuss future directions.\\

The starting point of our calculation is the Dirac-Coulomb Hamiltonian ($H_{\mathrm{DC}}$) given by

\begin{equation}
H_{\text{DC}} =  \sum_i \left [c \bm{\alpha}_i \cdot \bm{p}_i+(\bm{\beta_i}-1)c^2+V_n(\bm{r}_i)\right] 
 + \sum_{i > j}\frac{1}{r_{ij}} , \label{eq:HDC}
\end{equation}
where, $\bm{p}_i$ is the momentum for the $i$-th electron, $\bm{\alpha}_{i}$ and $\bm{\beta}_{i}$ are the Dirac matrix spinors, $c$ is the speed of light, $V_{\mathrm{n}}(\bm{r}_i)$ is the potential due to the nucleus and $r_{ij}$, is the interelectronic
separation. We generate the single-particle orbitals using a differential equation-based methodology within the MCDHF framework, in which atomic wavefunctions are expressed as linear combinations of configuration state functions (CSFs) denoted as $\Phi_r$ \cite{atoms11010007}. This MCDHF approach yields a reduced number of highly optimized single-particle orbitals, leading to a more compact CI Hamiltonian ($H_{\text{DC}}$) and a smaller set of expansion coefficients. The atomic state function (ASF), $|\Psi(JM_J\Pi)\rangle$, for a state with total angular momentum $J$, parity $\Pi$, and projection $M_J$ is expressed as
\begin{equation}
|\Psi(JM_J\Pi)\rangle = {\sum_{r=1}^{B}} c_r |\Phi_r(JM_J\Pi)\rangle,
\label{asf}
\end{equation}
where the $c_r$ are the expansion coefficients, {and $B$ is the number of CSFs}. 

For magnetic dipole hyperfine structure, we evaluate matrix elements of the operator
\begin{equation}
H_{\mathrm{hf}} = \sum_i \bm{\alpha}_{i} \cdot \frac{\bm{\mu}_I \times \bm{r}_i}{r_i^3},
\end{equation}
describing the interaction\cite{Das1973} between the electron current density and the vector
potential due to the nuclear magnetic dipole moment $\bm{\mu}_I$.




The HFS constant $A$ can be extracted directly in terms of the ASF expansion from Eq.~(\ref{asf}) and utilizing the Wigner-Eckart theorem.
\begin{equation}
A = A_{\text{fac}} \sum_{r,s} c_r^* c_s \langle \Phi_r(J \Pi) \| T_e^{(1)} \| \Phi_s(J \Pi) \rangle
\label{avalue}
\end{equation}
where  $ A_{\text{fac}} =\mu_I/(I\sqrt{J(J+1)(2J+1)})$ and $I$, $J$ are the total nuclear spin and electronic angular momenta, respectively. $\langle \Psi_r(J \Pi) \| T_e^{(1)} \| \Psi_s(J \Pi) \rangle$ is the reduced matrix element of the electronic operator, $T_e^{(1)}$ of the magnetic dipole hyperfine operator\cite{PhysRev.97.380,parpia1996grasp92}. 
    
The computational workflow separates quantum and classical components. Classical preprocessing using GRASP \cite{atoms11010007} generates two sets of data: (i) matrix elements $H_{rs} = \langle \Phi_{r} | H_{\mathrm{DC}} | \Phi_{s} \rangle$ defining the electronic structure problem, and (ii) reduced matrix elements $\langle \Phi_r || T_e^{(1)} || \Phi_s \rangle$ for hyperfine operator. The electronic structure eigenvalue problem is then reformulated as a quadratic unconstrained binary optimization (QUBO) problem suitable for quantum annealing hardware. The D-Wave annealer identifies the lowest-energy eigenstate by adiabatically evolving an initial transverse-field ground state into the final QUBO ground state, yielding the set of coefficients $\{c_r\}$ in the ASF basis and the corresponding energy. Next $A$ is computed by post-processing: quantum-derived   $\{c_r\}$ coefficients from the D-Wave annealer are contracted with classically computed operator matrix elements and $ A_{\text{fac}}$ according to Eq.~(\ref{avalue}).  The modified QAE framework is detailed below.

    \begin{figure*}[htbp!]
    
        \centering
        \includegraphics[width=0.95\textwidth]{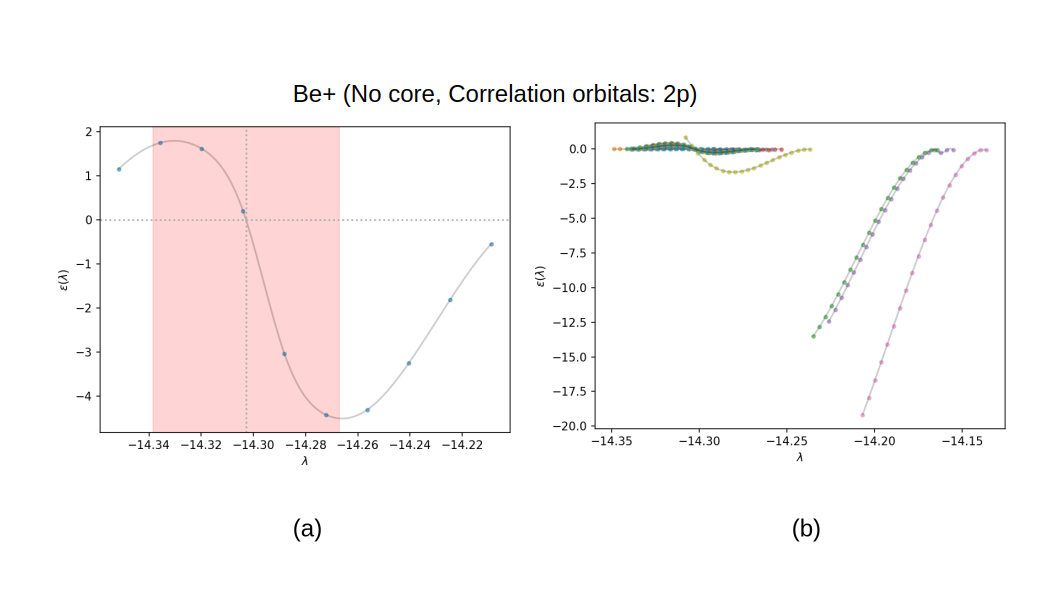}
        
        \caption{(a) The initial guess for $\lambda$ is obtained by a coarse scan of $\epsilon(\lambda)$ around {$H_{11}$} yielding the gray curve. The shaded pink region serves as the next narrowed down window, centered at the interpolated root of the functional, for the next iteration. (b) The subsequent scans over the narrower windows of $\lambda$ yield improved guesses for the Lagrange parameter.}
        
        \label{fig:estimatelambda}
        
    \end{figure*}


{The ground state solution of our modified QAE problem relies on finding the zero of the energy functional
\begin{equation}
    \epsilon(\lambda) = \braket{\Psi | H_{\text{DC}} | \Psi } - \lambda \braket{\Psi | \Psi }.
\end{equation}
With $\ket{\Psi}$ given in Eq.(\ref{asf})
the energy functional can be written in an intermediate quadratic form, suitable for D-Wave's quantum annealer hardware, as
\begin{equation}
    \epsilon(\lambda) = \sum_{r, s = 1}^{B}\limits c_{r} c_{s} H_{rs}- \lambda \sum_{r=1}^{B} c_{r}^2,
\end{equation}
where $H_{rs} = \braket{\Phi_r | H_{\text{DC}} | \Phi_s }$ are evaluated by  GRASP. The optimal value of the Lagrange multiplier $\lambda_\text{opt} \approx H_{11}$ corresponds to the ground state energy of the system.
}

$\epsilon(\lambda)$ is mapped onto a QUBO form using a floating-point encoding scheme, with each coefficient \( c_{r} \)  represented by \( K \)-bit binary variables, following the approach described in ~\cite{kumar2024computation}. The estimation of the optimal Lagrange multiplier ($\lambda_\text{opt}$) and \{$c_r$\} is done using zooming and sigma-annealing. Here sigma-annealing denotes the iterative convergence of $\{c_r\}$. Zooming refers to estimating $\lambda_\text{opt}$ by, first, an initial scan of $\lambda$ around {$H_{11}$} on a coarse grid 
(see Fig.~\ref{fig:estimatelambda}a), estimating $\lambda_\text{opt}$ by interpolation of the $\epsilon(\lambda)$ scan followed by iteratively scanning a narrower window of $\lambda$ centered at the previous zero of the energy functional. Such zooming yields a significant fraction of estimated roots around the actual $\lambda_\text{opt}$ with iteratively improving precision
(Fig.~\ref{fig:estimatelambda}b). Zooming on a coarse grid is key to accurately estimating $\lambda_\text{opt}$ during SA as well as during QA. Finally, the functional is evaluated on a finer grid in the last (and narrowest) window of $\lambda$ with the $\{ c_r \}$ taken from the grid point with the lowest magnitude of $\epsilon(\lambda)$. The plots in Fig.~\ref{fig:estimatelambda} are obtained from multiple SA runs for Be$^{+}$ (first case of Table \ref{tab:Be_results}). The final parameter window identified by SA is used for D-Wave's quantum annealer.


Under the QAE scheme, the  $B \times B$ Hamiltonian maps to a $BK \times BK$ matrix; we therefore restrict to $B=11$ for $K=10$ qubits for the full Hamiltonian. As already noted, a smaller $B$ reduces the cost of embedding on the annealer hardware which lacks all-to-all qubit connectivity.
For the number of CSFs exceeding 11, we had to resort to a truncated (and reordered) basis, estimating the contribution of the basis states by the magnitude of $c_r \approx H_{r 1}/(H_{11} - H_{rr})$, serving as the initial guess of the CSF coefficients from the independent electron pair approximation. This is also found to be in agreement with evaluations of $c_r$ from GRASP. The $r$-th row and column of the Hamiltonian were removed for all insignificant $c_r$. This worked well for all our chosen atomic systems (the last column of the Tables \ref{tab:Li_results},\ref{tab:Be_results}, \ref{tab:Na_results} and \ref{tab:Mg_results}) where $B$ was reduced from 14 to 11, 10, and 12, respectively.  Here 2p refers to $2p_{1/2}$ and $2p_{3/2}$ both and likewise for 3p.

\begin{table}[h!]
\caption{Ground-state energy and magnetic-dipole HFS constants of Li ($^2S_{1/2}$) computed with different correlation orbital sets. Here 2p refers to $2p_{1/2}$ and $2p_{3/2}$ both.Percentage deviations are defined as $\Delta(\%) = |Y_{\text{ref}} - Y_{\text{calc}}|/|Y_{\text{ref}}| \times 100$, where $Y$ represents the quantity (energy $E$ or HFS constant $A$), $Y_{\text{ref}}$ is the reference value, and $Y_{\text{calc}}$ is the calculated value. Reference values are GRASP results. Exc. denotes excitation type (S = single, D = double).}
\label{tab:Li_results}
\begin{ruledtabular}
\begin{tabular}{lrrr}
\toprule
Parameter & \multicolumn{3}{c}{Correlation orbitals} \\
\cmidrule(lr){2-4} & 2p & 3s & 2p,3s \\

Inactive core orbitals & None & None & None \\
Exc. & SD & SD & SD \\
No. of CSFs & 8 & 8 & $14 \rightarrow 11$ \\
\midrule
\multicolumn{4}{c}{\textit{Energy (a.u.)}} \\
GRASP & $-7.455$ & $-7.448$ & $-7.469$ \\
SA & $-7.453$ & $-7.448$ & $-7.468$ \\
$\Delta E_{\mathrm{SA}}(\%)$ & 0.027 & 0.000 & 0.013 \\
QA & $-7.456$ & $-7.448$ & $-7.468$ \\
$\Delta E_{\mathrm{QA}}(\%)$ & 0.013 & 0.000 & 0.013 \\
\midrule
\multicolumn{4}{c}{\textit{HFS constant $A$ (MHz)}} \\
GRASP & 285.938 & 392.935 & 387.726 \\
SA & 285.938 & 392.935 & 387.726 \\
$\Delta A_{\mathrm{SA}}(\%)$ & 0.000 & 0.000 & 0.000 \\
QA & 285.938 & 392.935 & 387.726 \\
$\Delta A_{\mathrm{QA}}(\%)$ & 0.000 & 0.000 & 0.000 \\
\bottomrule
\end{tabular}
\end{ruledtabular}
\end{table}


\begin{table}[h!]
\caption{Calculated ground-state energy and magnetic-dipole HFS constants for Be$^+$ ($^2S_{1/2}$) using GRASP, SA and QA with various correlation orbital sets.}
\label{tab:Be_results}
\begin{ruledtabular}
\begin{tabular}{lrrr}
\toprule
Parameter & \multicolumn{3}{c}{Correlation orbitals} \\
\cmidrule(lr){2-4} & 2p & 3s & 2p,3s \\
Inactive core orbitals & None & None & None \\
Exc. & SD & SD & SD \\
No. of CSFs & 8 & 8 & $14 \rightarrow 11$ \\
\midrule
\multicolumn{4}{c}{\textit{Energy (a.u.)}} \\
GRASP & $-14.280$ & $-14.294$ & $-14.317$ \\
SA & $-14.304$ & $-14.295$ & $-14.316$ \\
$\Delta E_{\mathrm{SA}}(\%)$ & 0.168 & 0.007 & 0.007 \\
QA & $-14.302$ & $-14.265$ & $-14.316$ \\
$\Delta E_{\mathrm{QA}}(\%)$ & 0.154 & 0.203 & 0.007 \\
\midrule
\multicolumn{4}{c}{\textit{HFS constant $A$ (MHz)}} \\
GRASP & $-505.199$ & $-622.993$ & $-619.058$ \\
SA & $-505.199$ & $-622.993$ & $-619.058$ \\
$\Delta A_{\mathrm{SA}}(\%)$ & 0.000 & 0.000 & 0.000 \\
QA & $-505.199$ & $-622.993$ & $-619.058$ \\
$\Delta A_{\mathrm{QA}}(\%)$ & 0.000 & 0.000 & 0.000 \\
\bottomrule
\end{tabular}
\end{ruledtabular}
\end{table}

\begin{figure}[h!]
    \centering
    \includegraphics[width=0.95\linewidth]{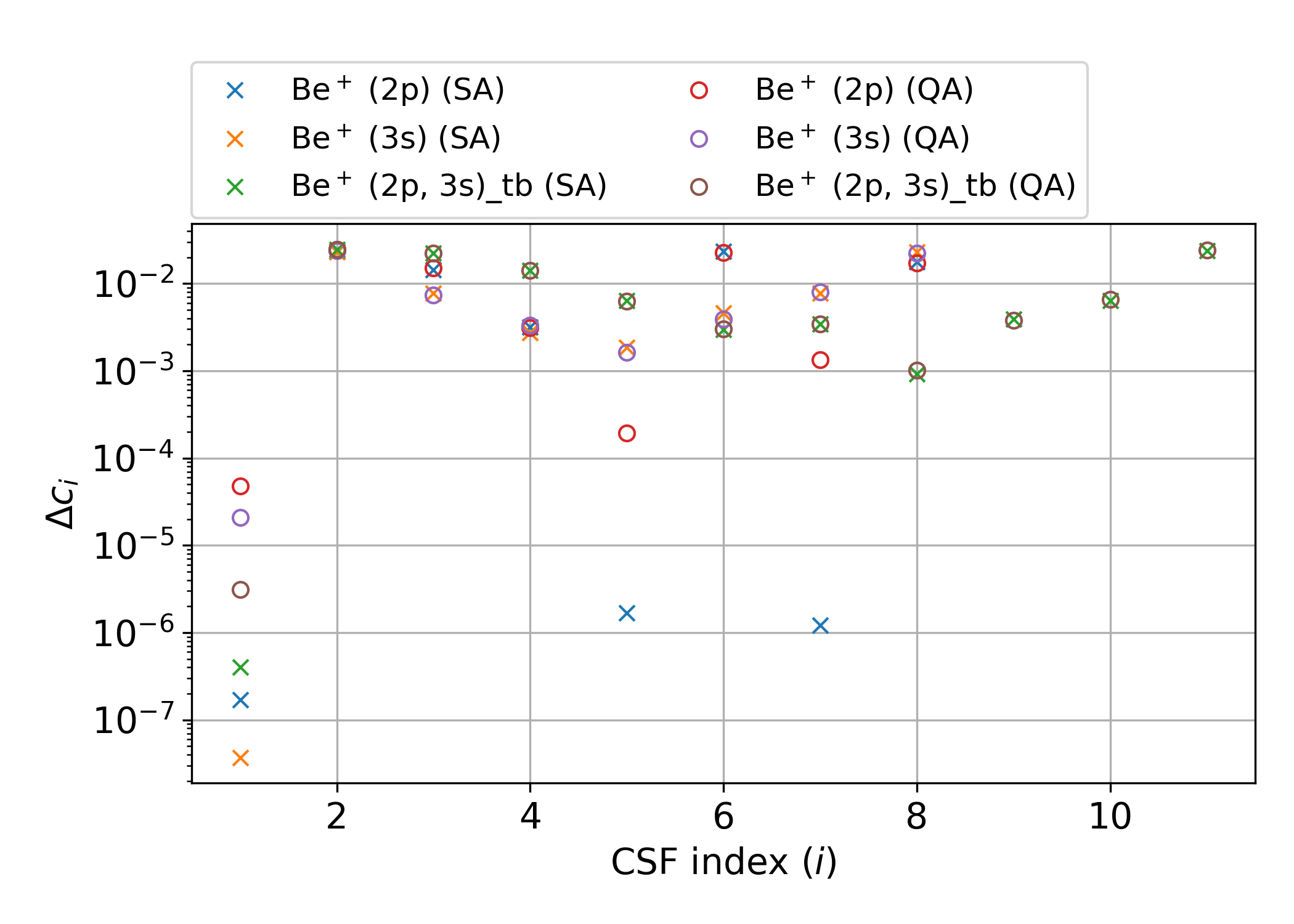}
    \caption{The absolute error in coefficients computed by SA and QA with respect to GRASP for Be$^+$ for the cases listed in Table \ref{tab:Be_results}. The list of active orbitals are in the parenthesis following the element name. {\tt tb} in the legends refer to a truncated basis.}
    \label{fig:cierrBe}
\end{figure}

\begin{table}[h!]
\caption{Ground-state energy and HFS constants for Na ($^2S_{1/2}$) across different correlation orbital configurations.  Here 3p refers to $3p_{1/2}$ and $3p_{3/2}$ both. }
\label{tab:Na_results}
\begin{ruledtabular}
\begin{tabular}{lrrrr}
\toprule
Parameter & \multicolumn{4}{c}{Correlation orbitals} \\
\cmidrule(lr){2-5}
 & 3p (S) & 4s (SD) & 3p,4s (S) & 4s (SD) \\
\midrule
Inactive core orbitals & 1s,2s & 1s,2s & 1s & 1s \\
No. of CSFs & 7 & 8 & 11 & $14 \rightarrow10$ \\
\midrule
\multicolumn{5}{c}{\textit{Energy (a.u.)}} \\
GRASP & $-162.078$ & $-162.080$ & $-162.078$ & $-162.083$ \\
SA & $-162.100$ & $-162.100$ & $-162.100$ & $-162.100$ \\
$\Delta E_{\mathrm{SA}}(\%)$ & 0.014 & 0.012 & 0.014 & 0.010 \\
QA & $-162.100$ & $-162.100$ & $-162.100$ & $-162.100$ \\
$\Delta E_{\mathrm{QA}}(\%)$ & 0.014 & 0.012 & 0.014 & 0.010 \\
\midrule
\multicolumn{5}{c}{\textit{HFS constant $A$ (MHz)}} \\
GRASP & 629.649 & 627.758 & 719.876 & 736.080 \\
SA & 629.649 & 627.758 & 719.876 & 736.148 \\
$\Delta A_{\mathrm{SA}}(\%)$ & 0.000 & 0.000 & 0.000 & 0.009 \\
QA & 629.649 & 627.758 & 719.876 & 736.148 \\
$\Delta A_{\mathrm{QA}}(\%)$ & 0.000 & 0.000 & 0.000 & 0.009 \\
\bottomrule
\end{tabular}
\end{ruledtabular}
\end{table}


\begin{table}[h!]
\caption{Ground-state energy and HFS constants of Mg$^+$ ($^2S_{1/2}$) for varying correlation orbital sets.}
\label{tab:Mg_results}
\begin{ruledtabular}
\begin{tabular}{lrrrr}
\toprule
Parameter & \multicolumn{4}{c}{Correlation orbitals} \\
\cmidrule(lr){2-5}
 & 3p (S) & 4s (SD) & 3p,4s (S) & 4s (SD) \\
\midrule
Inactive core orbitals & 1s,2s & 1s,2s & 1s & 1s \\
No. of CSFs & 7 & 8 & 11 & $14 \rightarrow 12$ \\
\midrule
\multicolumn{5}{c}{\textit{Energy (a.u.)}} \\
GRASP & $-199.692$ & $-199.694$ & $-199.692$ & $-199.696$ \\
SA & $-199.700$ & $-199.727$ & $-199.700$ & $-199.727$ \\
$\Delta E_{\mathrm{SA}}(\%)$ & 0.004 & 0.017 & 0.004 & 0.016 \\
QA & $-199.700$ & $-199.727$ & $-199.700$ & $-199.727$ \\
$\Delta E_{\mathrm{QA}}(\%)$ & 0.004 & 0.017 & 0.004 & 0.016 \\
\midrule
\multicolumn{5}{c}{\textit{HFS constant $A$ (MHz)}} \\
GRASP & $-469.872$ & $-469.329$ & $-523.903$ & $-531.086$ \\
SA & $-469.872$ & $-469.329$ & $-523.903$ & $-531.086$ \\
$\Delta A_{\mathrm{SA}}(\%)$ & 0.000 & 0.000 & 0.000 & 0.000 \\
QA & $-469.872$ & $-469.329$ & $-523.903$ & $-531.086$ \\
$\Delta A_{\mathrm{QA}}(\%)$ & 0.000 & 0.000 & 0.000 & 0.000 \\
\bottomrule
\end{tabular}
\end{ruledtabular}
\end{table}
\vspace{0.2cm}


The computed ground-state energies and magnetic-dipole HFS constants for Li, Be$^{+}$, Na, and Mg$^{+}$ are presented in Tables~\ref{tab:Li_results}--\ref{tab:Mg_results}, with quantum computations performed on the D-Wave Advantage system 6.4 quantum processing unit (QPU) based on Pegasus topology \cite{ PhysRevApplied.21.034023,boothby2020nextgenerationtopologydwavequantum} with over 5000 qubits, utilizing 1000 reads and default chain coupling strength\cite{dwave_chain_strength_2022} using D-Wave’s Ocean SDK \cite{ocean_sdk_2023}. For our modified QAE calculations, we restricted the correlation orbitals to 3s for Li and Be$^+$ and 4s for Na and Mg$^+$, with a small number of CSFs due to the present challenges of increased encoding complexity and subtle convergence issues on the D-Wave quantum annealer. The absolute errors in $\{c_{r}\}$ relative to GRASP for Be$^{+}$ 
(Table~\ref{tab:Be_results}) are shown in Fig.~\ref{fig:cierrBe} 
as the worst-case scenario among the cases considered, the figures for the remaining 
three atomic cases are provided in the supplemental material.  The error for $c_1$ in each case is the lowest ($< 10^{-4}$). The remarkably close errors for SA and QA in the subsequent $c_{r}$ are attributed to the effectiveness of the zooming strategy to estimate $\lambda_\text{opt}$. The QA-computed $|c_1|> |c_{r(>1)}|$  (tabulated in the supplemental material) will have a significantly small contribution to HFS constants (by virtue of the quadratic dependence of $A$ on $\{ c_{r} \}$ Eq. (\ref{avalue})) and therefore the larger absolute errors for $r>1$ have insignificant effect on $\Delta A_{QA}$. Sigma-annealing, combined with zooming over a recentered window, enhances the precision of the ground-state eigenvector estimate.

For the ground state of Li ($^2S_{1/2}$), both SA and QA reproduce GRASP energies with remarkable accuracy, yielding relative deviations below 0.03\% across all correlation orbital configurations, while HFS constants remain identical across all three methods within the chosen numerical precision. The exceptional agreement stems from careful consideration of the underlying physics governing hyperfine interactions and the iterative zooming and sigma annealing strategy in our modified QAE for ground-state energy and eigenvector estimation.

The radial probability distribution of s-orbitals exhibits significant amplitude in the nuclear region, unlike p-, d-, or higher angular momentum orbitals which nearly vanishes at the nucleus, thus inclusion of virtual s-orbitals in relatively small CSF bases (12 or fewer 
{ CSFs}) efficiently captures essential electron correlation effects, with the 3s correlation orbital contributing significantly compared to 2p alone, demonstrating that s-orbital correlation dominates hyperfine interactions in light alkali-like systems. Similar trends emerge for the ground state of Be$^{+}$ ($^2S_{1/2}$) in Table~\ref{tab:Be_results}, where GRASP and annealing-based ground state eigenvectors show a significant difference of the HFS constants upon inclusion of s-orbitals with higher principal quantum number. 

For heavier systems, the ground state of Na ($^2S_{1/2}$) and Mg$^{+}$ ($^2S_{1/2}$), presented in Tables~\ref{tab:Na_results} and \ref{tab:Mg_results}, GRASP and annealing energies remain consistent within 0.02\% across different correlation orbital sets, confirming the robustness of our optimization approach. However, the HFS constant exhibits much more sensitivity  to the correlation orbitals as well as to the inactive core orbitals. As the nuclear charge $Z$ increases, the nuclear charge distribution extends over larger spatial regions, modifying electron-nucleus interactions and enhancing relativistic effects, necessitating a comprehensive treatment of electron correlation that incorporates virtual orbitals of type p$_{1/2}$ along with orbitals of type s to achieve high accuracy.


In summary, we have computed the ground state energies and magnetic dipole hyperfine constants of $^{7}\mathrm{Li}$, $^{9}\mathrm{Be}^{+}$, $^{23}\mathrm{Na}$, and $^{25}\mathrm{Mg}^{+}$ using the modified QAE algorithm and the CI method with MCDHF orbitals. The careful inclusion of correlation orbitals with significant contributions to hyperfine interactions, together with our iterative annealing strategy in the modified QAE framework, yields accurate estimates of expansion coefficients, ground state energies and HFS constants that closely agree with classical CI and SA results up to three decimal places accuracy for HFS constants across the different systems considered. Our proposed framework and findings indicate that it can be readily extended to the calculation of other atomic properties, in particular electric-quadrupole HFS constants and isotope-shift parameters. Future possible advances in quantum annealing hardware, such as the D-Wave Advantage2 QPU with Zephyr topology\cite{Boothby2022Zephyr} and improved solvers\cite{osaba2024dwavesnonlinearprogramhybridsolver}, are expected to reduce embedding overheads and enhance solution accuracy. Continued developments in both hardware and algorithms, including improved preprocessing and error-mitigation techniques \cite{yip2025variationalquantumannealingquantum,Raymond2025,Djidjev2024ReplicationBasedQA}, may enable quantum annealers to play an increasingly important role in the accurate determination of atomic and molecular properties.

\textit{Acknowledgment:} BP thanks Dr. Ranjan Modak (IIT Tirupati) for his continuous support throughout this project. SD and BP thank Pradyot Pritam Sahoo and Dr. Arup Chakraborty for useful discussions. PJ acknowledges support from the Swedish Research Council (VR 2023-05367).\\
\textit{Data Availablity Statement:} The data are available from the authors upon reasonable request.

%

\end{document}


\title{Supplemental
 Materials}
\maketitle

\newpage 
\listoftables
\listoffigures
\newpage 

\section{CSF coefficient tables}
The CSF coefficients are computed using GRASP, simulated annealing (SA) and by quantum annealing (QA) on D-Wave's annealing hardware listed in the following tables.

\begin{table}[h!]
    \centering
    \begin{tabular}{cccc}\hline
         CSF index&  GRASP&  SA& QA\\\hline
         1&  -9.99137E-01&  9.99E-01& 9.99E-01\\
         2&  1.04839E-03&  3.38E-02& 3.38E-02\\
         3&  3.26301E-03&  2.39E-02& 2.39E-02\\
         4&  -9.13394E-07&  -3.26E-03& -3.22E-03\\
         5&  2.30606E-03&  -2.31E-03& -2.32E-03\\
         6&  -3.37513E-02&  -1.05E-03& -1.01E-03\\
         7&  9.02786E-09&  1.15E-08& -4.44E-05\\
         8&  -2.38626E-02&  8.93E-07& -1.33E-05\\\hline
    \end{tabular}
    \caption{CSF coefficients for Li with 2p as correlation orbital. }
    \label{tab:csfcoeffs_Li_2p}
\end{table}

\begin{table}[h!]
    \centering
    \begin{tabular}{cccc}\hline
         CSF index&  GRASP&  SA& QA\\\hline
         1&  -9.99327E-01&  -9.99E-01& 9.99E-01\\
         2&  -3.09311E-04&  3.47E-02& -3.48E-02\\
         3&  -9.24112E-04&  -8.90E-03& 8.84E-03\\
         4&  2.63982E-03&  -6.54E-03& 6.55E-03\\
         5&  -1.53516E-03&  2.59E-03& -2.56E-03\\
         6&  -6.56503E-03&  -1.55E-03& 1.59E-03\\
         7&  -8.85121E-03&  -9.14E-04& 9.06E-04\\
         8&  3.48274E-02&  -3.57E-04& 2.76E-04\\\hline
    \end{tabular}
    \caption{CSF coefficients for Li with 3s as correlation orbital.  }
    \label{tab:csfcoeffs_Li_3s}
\end{table}


\begin{table}[h!]
    \centering
    \begin{tabular}{cccc}\hline
         CSF index&  GRASP&  SA& QA
\\\hline
         1&  -9.98508E-01&  -9.99E-01& -9.98E-01
\\
         2&  -3.32758E-04&  -3.32E-04& -1.21E-04
\\
         3&  -9.89504E-05&  & 
\\
         4&  2.74099E-03&  2.74E-03& 2.89E-03
\\
         5&  -1.23020E-03&  -1.23E-03& -1.06E-03
\\
         6&  3.20031E-03&  3.20E-03& 3.03E-03
\\
         7&  -8.78882E-07&  & 
\\
 8& 2.26178E-03& 2.26E-03&2.29E-03
\\
 9& -6.47866E-03& -6.48E-03&-6.66E-03
\\
 10& -8.66571E-03& -8.66E-03&-8.42E-03
\\
         11&  -3.33577E-02&  -3.34E-02& -3.36E-02
\\
 12& 2.78875E-08& &
\\
 13& -2.35841E-02& -2.36E-02&-2.36E-02
\\
 14& 3.42169E-02& 3.42E-02&3.43E-02\\\hline
    \end{tabular}
    \caption{CSF coefficients for Li with 2p and 3s as correlation orbitals. For the SA and QA calculations, the basis was truncated from 14 to 11 CSFs. The truncated coefficients have been left blank in the columns for SA and QA in this table and subsequent tables with a truncated basis. }
    \label{tab:csfcoeffs_Li_2p3s}
\end{table}

\begin{table}[h!]
    \centering
    \begin{tabular}{cccc}\hline
         CSF index&  GRASP&  SA& QA\\\hline
         1&  -9.99523E-01&  1.00E+00& -9.99E-01\\
         2&  1.57928E-03&  2.50E-02& -2.63E-02\\
         3&  3.18614E-03&  1.77E-02& -1.83E-02\\
         4&  -1.98478E-06&  -3.19E-03& 3.11E-03\\
         5&  2.25036E-03&  -2.25E-03& 2.44E-03\\
         6&  -2.49809E-02&  -1.58E-03& 2.40E-03\\
         7&  1.69440E-08&  1.24E-06& 1.34E-03\\
 8& -1.76597E-02& -1.81E-06&3.73E-04\\\hline
    \end{tabular}
    \caption{CSF coefficients for Be$^{+}$ with 2p as correlation orbitals.  }
    \label{tab:csfcoeffs_Be_2p}
\end{table}

\begin{table}[h!]
    \centering
    \begin{tabular}{cccc}\hline
         CSF index&  GRASP&  SA& QA\\\hline
         1&  9.99668E-01&  -1.00E+00& 1.00E+00\\
         2&  2.42124E-04&  2.34E-02& -2.42E-02\\
         3&  -2.55474E-04&  -7.99E-03& 7.59E-03\\
         4&  -3.43566E-03&  -6.17E-03& 6.77E-03\\
         5&  1.59038E-03&  3.44E-03& -3.22E-03\\
         6&  6.18062E-03&  -1.59E-03& 2.24E-03\\
         7&  7.98807E-03&  2.58E-04& 2.25E-05\\
 8& -2.34018E-02& -2.46E-04&9.46E-04\\\hline
    \end{tabular}
    \caption{CSF coefficients for Be$^{+}$ with 3s as correlation orbitals. }
    \label{tab:csfcoeffs_Be_3s}
\end{table}


\begin{table}[h!]
    \centering
    \begin{tabular}{cccc}\hline
         CSF index&  GRASP&  SA& QA\\\hline
         1&  -9.99207E-01&  -9.99E-01& 9.99E-01
\\
         2&  -2.52669E-04&  & 

\\
         3&  8.03884E-04&  8.06E-04& -6.94E-04
\\
         4&  3.45427E-03&  3.45E-03& -3.44E-03
\\
         5&  -1.47029E-03&  -1.47E-03& 1.33E-03
\\
         6&  3.13996E-03&  3.14E-03& -3.23E-03
\\
         7&  -1.93199E-06&  & 

\\
 8& 2.21778E-03& 2.22E-03&-2.35E-03
\\
 9& -6.12820E-03& -6.12E-03&6.14E-03
\\
 10& -7.88207E-03& -7.88E-03&7.76E-03
\\
 11& -2.47845E-02& -2.48E-02&2.49E-02
\\
 12& 5.03620E-08& &

\\
 13& -1.75206E-02& -1.75E-02&1.76E-02
\\
 14& 2.31127E-02& 2.31E-02&-2.31E-02\\\hline
    \end{tabular}
    \caption{CSF coefficients for Be$^{+}$ with 2p and 3s as correlation orbitals. For the SA and QA approaches, a truncated basis of 11 CSFs was used in place of the full 14 CSFs basis. }
    \label{tab:csfcoeffs_Be_2p3s}
\end{table}

\begin{table}[h!]
    \centering
    \begin{tabular}{cccc}\hline
         CSF index&  GRASP&  SA& QA\\\hline
         1&  -9.99947E-01&  -1.00E+00& -1.00E+00\\
 2& 4.69660E-03& -6.83E-03&-6.83E-03\\
 3& -1.67244E-04& 4.69E-03&4.69E-03\\
         4&  -3.60695E-03&  4.62E-03& 4.62E-03\\
         5&  -6.93694E-03&  -3.55E-03& -3.55E-03\\
         6&  -2.52152E-04&  -2.51E-04& -2.49E-04\\
         7&  4.76181E-03&  -1.66E-04& -1.65E-04\\\hline
    \end{tabular}
    \caption{CSF coefficients for Na with 3p as correlation orbitals, single excitations and inactive 2s orbital.  }
    \label{tab:csfcoeffs_Na_3s_S_in_2s}
\end{table}

\begin{table}[h!]
    \centering
    \begin{tabular}{cccc}\hline
         CSF index&  GRASP&  SA& QA\\\hline
         1&  -9.99853E-01&  -1.00E+00& 1.00E+00\\
 2& 1.26769E-04& -1.35E-02&1.35E-02\\
 3& 2.55882E-03& -9.53E-03&9.53E-03\\
         4&  -9.55508E-03&  3.61E-03& -3.61E-03\\
         5&  -3.79618E-06&  2.55E-03& -2.55E-03\\
         6&  7.91169E-08&  1.25E-04& -1.32E-04\\
         7&  3.62670E-03&  -3.77E-06& 4.37E-06\\
 8& -1.35086E-02& 2.10E-08&-1.37E-06\\\hline
    \end{tabular}
    \caption{CSF coefficients for Na with 4s as correlation orbital, single and double excitations and inactive 2s orbital.  }
    \label{tab:csfcoeffs_Na_4s_SD_in_2s}
\end{table}

\begin{table}[h!]
    \centering
    \begin{tabular}{cccc}\hline
         CSF index&  GRASP&  SA& QA\\\hline
         1&  9.99942E-01&  -1.00E+00& 1.00E+00\\
 2& -1.47100E-04& -7.16E-03&7.17E-03\\
 3& -4.63802E-03& 4.64E-03&-4.64E-03\\
         4&  3.48711E-04&  4.57E-03& -4.57E-03\\
         5&  3.88697E-03&  -3.83E-03& 3.83E-03\\
         6&  7.27083E-03&  -1.76E-03& 1.76E-03\\
         7&  7.90288E-04&  -7.81E-04& 7.69E-04\\
 8& -4.71327E-03& 6.35E-04&-6.32E-04\\
 9& 1.05072E-04& -3.45E-04&3.43E-04\\
 10& 1.76770E-03& 1.44E-04&-1.48E-04\\
 11& -6.34766E-04& -1.07E-04&1.07E-04\\\hline
    \end{tabular}
    \caption{CSF coefficients for Na with 3p and 4s as correlation orbitals, single excitations and inactive 1s orbital. }
    \label{tab:csfcoeffs_Na_3p4s_S_in_1s}
\end{table}


\begin{table}[h!]
    \centering
    \begin{tabular}{cccc}\hline
         CSF index&  GRASP&  SA& QA\\\hline
         1&  -9.99744E-01&  -1.00E+00& 1.00E+00
\\
 2& 5.69228E-06& &
\\
 3& 2.21598E-03& 2.20E-03&-2.20E-03
\\
         4&  -8.63089E-03&  -8.61E-03& 8.61E-03
\\
         5&  -3.73102E-06&  & 
\\
         6&  7.93700E-08&  & 
\\
         7&  3.13917E-03&  3.12E-03& -3.12E-03
\\
 8& -1.21952E-02& -1.22E-02&1.22E-02
\\
 9& -6.22390E-04& &
\\
 10& 1.21128E-03& 1.20E-03&-1.19E-03
\\
 11& -1.02385E-03& -1.01E-03&1.01E-03
\\
 12& -3.19135E-03& -3.19E-03&3.19E-03
\\
 13& -3.85343E-03& -3.83E-03&3.83E-03
\\
 14& 1.57030E-02& 1.57E-02&-1.57E-02\\\hline
    \end{tabular}
    \caption{CSF coefficients for Na with 4s as correlation orbital, single and double excitations and inactive 1s orbital. In the SA and QA calculations, the basis set was reduced from 14 to 11 configurations.  }
    \label{tab:csfcoeffs_Na_4s_SD_in_1s}
\end{table}

\begin{table}[h!]
    \centering
    \begin{tabular}{cccc}\hline
         CSF index&  GRASP&  SA& QA\\\hline
         1&  -9.99945E-01&  -1.00E+00& -1.00E+00\\
 2& 4.74811E-03& -7.06E-03&-7.06E-03\\
 3& -1.49907E-04& 4.82E-03&4.82E-03\\
         4&  -3.62236E-03&  4.73E-03& 4.73E-03\\
         5&  -7.08915E-03&  -3.61E-03& -3.61E-03\\
         6&  -3.38953E-04&  -3.38E-04& -3.43E-04\\
         7&  4.83468E-03&  -1.49E-04& -1.52E-04\\\hline
    \end{tabular}
    \caption{CSF coefficients for Mg$^{+}$ with 3p as correlation orbital, single excitations and inactive 2s orbital. }
    \label{tab:csfcoeffs_Mg_3p_S_in_2s}
\end{table}

\begin{table}[h!]
    \centering
    \begin{tabular}{cccc}\hline
         CSF index&  GRASP&  SA& QA\\\hline
         1&  -9.99910E-01&  -1.00E+00& -1.00E+00\\
 2& 1.16706E-04& -1.02E-02&-1.02E-02\\
 3& 2.69553E-03& -7.22E-03&-7.22E-03\\
         4&  -7.24518E-03&  3.80E-03& 3.80E-03\\
         5&  -4.28798E-06&  2.68E-03& 2.68E-03\\
         6&  8.42385E-08&  1.16E-04& 1.16E-04\\
 7& 3.81711E-03& -4.30E-06&-3.57E-06\\
 8& -1.02372E-02& 8.71E-08&1.54E-06\\\hline
    \end{tabular}
    \caption{CSF coefficients for Mg$^{+}$ with 4s as correlation orbital, single and double excitations and inactive 2s orbital.  }
    \label{tab:csfcoeffs_Mg_4s_SD_in_2s}
\end{table}

\begin{table}[h!]
    \centering
    \begin{tabular}{cccc}\hline
         CSF index&  GRASP&  SA& QA\\\hline
         1&  -9.99940E-01&  -1.00E+00& 1.00E+00\\
 2& 1.87040E-04& -7.02E-03&7.02E-03\\
 3& 4.68714E-03& 4.76E-03&-4.75E-03\\
         4&  -1.74490E-04&  4.67E-03& -4.68E-03\\
         5&  -3.62178E-03&  -3.61E-03& 3.62E-03\\
         6&  -7.04531E-03&  -3.24E-03& 3.24E-03\\
 7& -3.94901E-04& 9.21E-04&-9.16E-04\\
 8& 4.77297E-03& -3.93E-04&3.88E-04\\
 9& -7.91008E-05& 1.86E-04&-1.98E-04\\
 10& -3.23978E-03& -1.74E-04&1.77E-04\\
 11& 9.22300E-04& -7.99E-05&8.29E-05\\\hline
    \end{tabular}
    \caption{CSF coefficients for Mg$^{+}$ with 3p, 4s as correlation orbital, single excitations and inactive 1s orbital.   }
    \label{tab:csfcoeffs_Mg_3p4s_S_in_1s}
\end{table}


\begin{table}[h!]
    \centering
    \begin{tabular}{cccc}\hline
         CSF index&  GRASP&  SA& QA\\\hline
         1&  -9.99823E-01&  -1.00E+00& 1.00E+00
\\
 2& -2.11337E-05& -2.07E-05&2.08E-05
\\
 3& 2.44065E-03& 2.43E-03&-2.43E-03
\\
         4&  -6.76920E-03&  -6.75E-03& 6.75E-03
\\
         5&  -4.24704E-06&  & 
\\
         6&  9.76674E-08&  & 
\\
 7& 3.45502E-03& 3.44E-03&-3.44E-03
\\
 8& -9.56096E-03& -9.54E-03&9.54E-03
\\
 9& -5.77374E-04& -5.71E-04&5.74E-04
\\
 10& 2.48906E-03& 2.48E-03&-2.48E-03
\\
 11& -1.08582E-03& -1.08E-03&1.08E-03
\\
 12& -3.83242E-03& -3.82E-03&3.82E-03
\\
 13& -4.50839E-03& -4.50E-03&4.50E-03
\\
 14& 1.25179E-02& 1.25E-02&-1.25E-02\\\hline
    \end{tabular}
    \caption{CSF coefficients for Mg$^{+}$ with 4s as correlation orbital, single and double excitations and inactive 1s orbital. For the SA and QA approaches, a truncated basis of 12 configurations was used in place of the full 14 configuration basis.  }
    \label{tab:csfcoeffs_Mg_4s_SD_in_1s}
\end{table}

\clearpage
\section{The absolute error in CSF coefficients}

\begin{figure}[h!]
    \centering
    \includegraphics[width=0.95\linewidth]{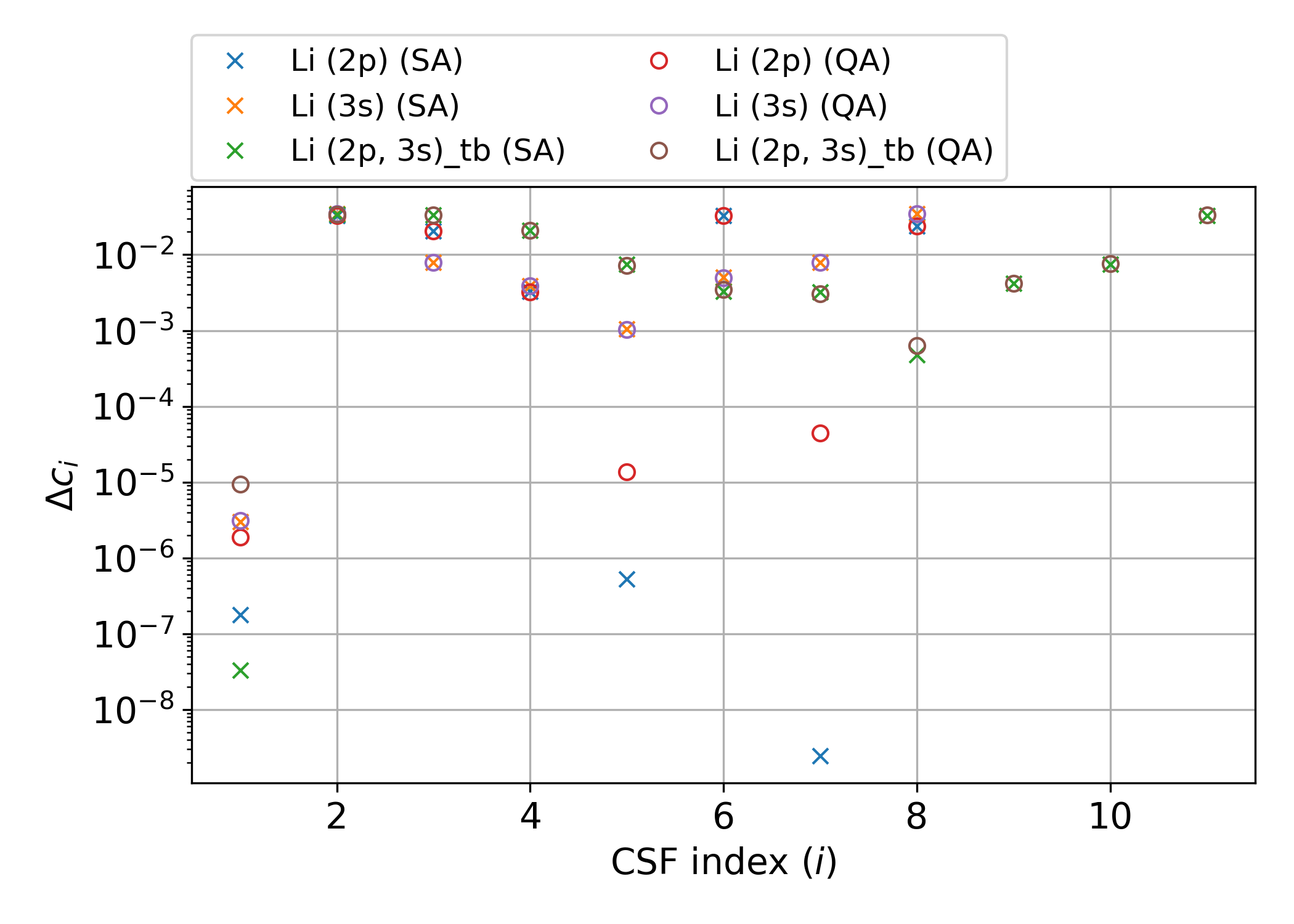}
    \caption{The absolute error in coefficients computed by SA and QA with respect to GRASP for Li for the cases listed in Table I in the main manuscript. The list of active orbitals are in the parenthesis following the element name. {\tt tb} in the legends refer to a truncated basis.}
    \label{fig:cierrLi}
\end{figure}

\begin{figure}[h!]
    \centering
    \includegraphics[width=0.95\linewidth]{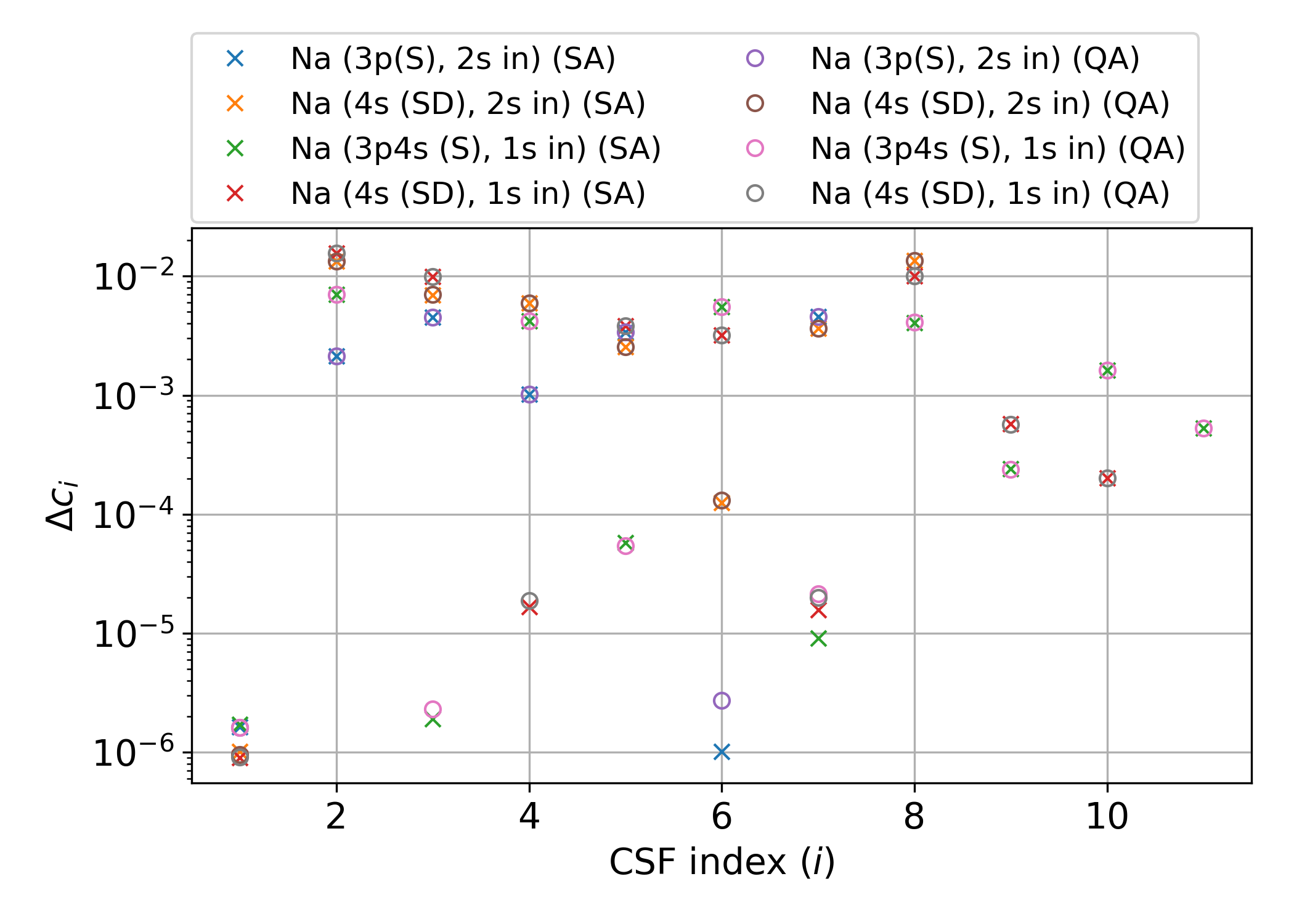}
    \caption{The absolute error in coefficients computed by SA and QA with respect to GRASP for Na for the cases listed in Table III in the main manuscript. S and SD refer to single and double excitations. {\tt in} in the legend refers to inactive orbitals. The other legends are identical to Fig. \ref{fig:cierrLi}.}
    \label{fig:cierrNa}
\end{figure}

\begin{figure}[h!]
    \centering
    \includegraphics[width=0.95\linewidth]{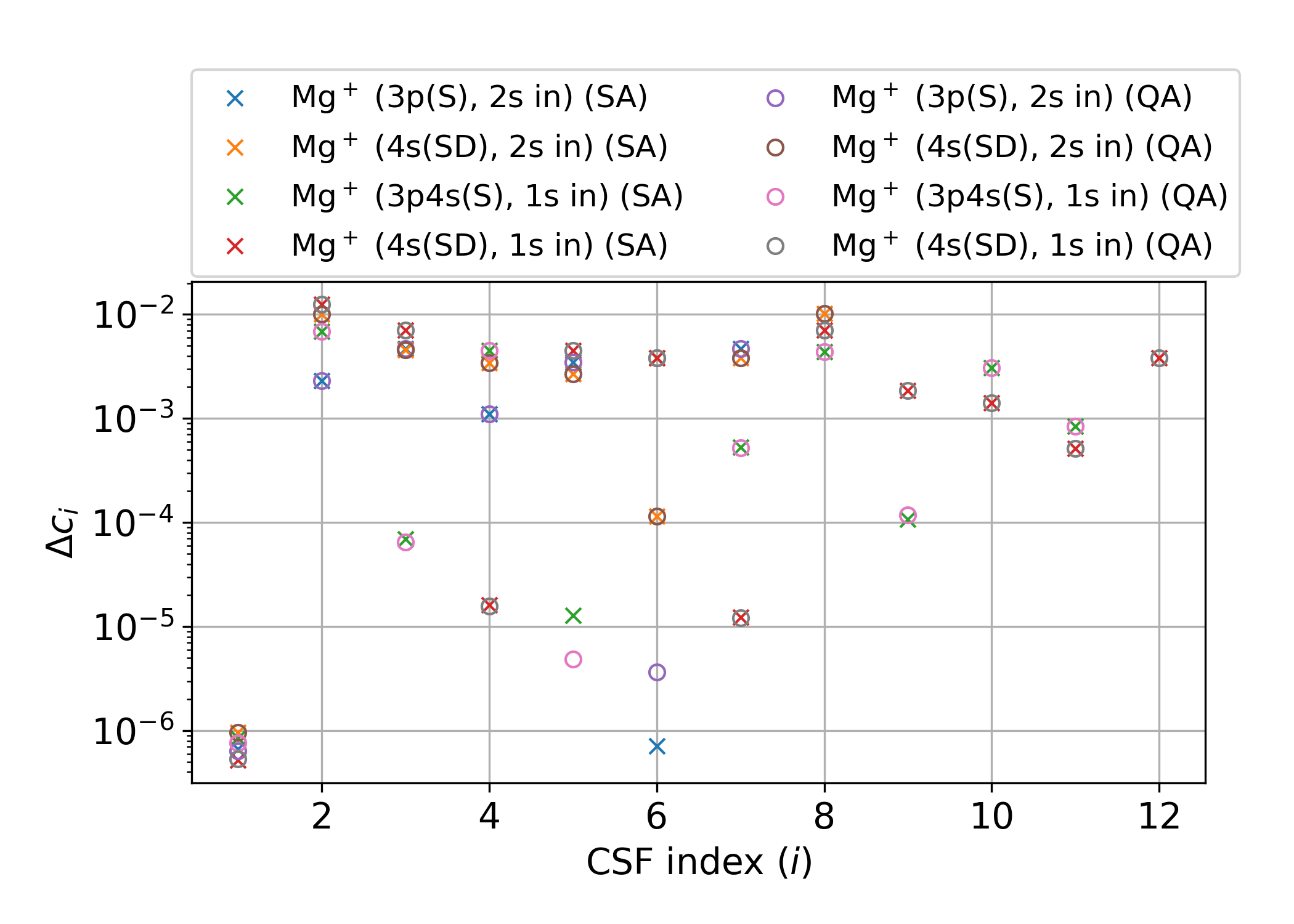}
    \caption{The absolute error in coefficients computed by SA and QA with respect to GRASP for Mg$^+$ for the cases listed in Table IV in the main manuscript. The legends are identical to Fig. \ref{fig:cierrLi}.}
    \label{fig:cierrMg}
\end{figure}